# General solution for quantitative dark-field contrast imaging with grating interferometers

M. Strobl[1]

[1] European Spallation Source ESS AB, Instrument division, Tunavaegan 24, 22100 Lund, Sweden

Abstract
Dark-field contrast imaging with grating interferometers has proven to hold huge potential for numerous applications with X-rays and with neutrons conveying biology and medicine as well as engineering and magnetism, respectively. However, a concept to extract quantitative information is still missing. Here a general theory as well as a measurement strategy is introduced, allowing extraction of quantitative small-angle scattering information such as structure sizes and scattering cross sections. The validity of the description is demonstrated by a specific example from literature.

I. INTRODUCTION

Over the past decade – since the introduction of grating interferometers [1-4] (Fig. 1) for imaging – corresponding dark-field imaging with X-rays [5] as well as with neutrons [6] has turned out as a remarkable success [7-19] and is advancing to even being introduced to diagnostic medical imaging with X-rays [7-10] in the near future. In imaging with a grating interferometer a sinusoidal modulation function is measured in every pixel of an image by scanning an absorption grating to achieve sub-pixel resolution of the corresponding interference pattern induced by a phase grating. This way not only the conventional attenuation image, which corresponds to the mean intensity of this pattern in a pixel, but also a differential phase contrast image and a dark-field image can be generated from the data. The phase contrast is measured by the relative phase shift of the modulation pattern and allows for mapping the refractive index distribution in addition to the attenuation image, providing complementary image contrast in many cases. The dark-field signal is based on the relative visibility, i.e. modulation amplitude and is related to small angle scattering induced by the sample by sub-resolution structure sizes. The dark-field signal hence provides access to structural information beyond spatial resolution of the imaging instrument.

Applications of dark-field contrast with neutrons up to now concentrate mainly on engineering materials [14] but especially also magnetic materials [15-19], where neutrons due to their magnetic moment provide remarkable contrast. For X-ray dark-field contrast imaging the ability of this technique to access soft tissue early on moved the focus to medical applications in particular also with regards to cancer diagnostics and research [8-10].

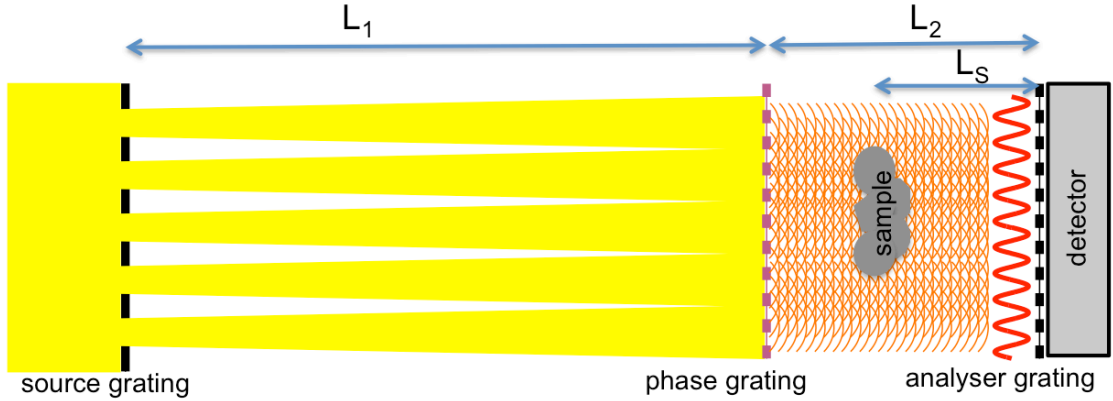

FIG. 1 Schematic sketch of grating interferometer set-up including three gratings, the corresponding distances and sample area referred to in this work; the cosine modulation function at the analyzer grating right in front of the detector is sketched as well.

This success of the method, even though still based on qualitative results, besides clinical tests for medical applications [8-10], also triggered a remarkable effort in fully understanding and describing the measured signals theoretically in order to finally extract quantitative information [12,20-29]. However, up to date these efforts have either been limited to establish relations between the signal measured and the thickness [6,11,12,20-24,26,29], which in turn allows establishing and utilizing conventional tomographic reconstruction, i.e. retrieving the corresponding signal in 3D, or to solutions for very specific scattering structures with the potential to describe the signal when the sample parameters are known in detail beforehand [27,28]. The former attempts [11,12,20-24,26,29] in general re-establish the approximation already published a decade ago [30,31] and used for first tomographic reconstructions of the dark-field signal from a grating interferometer [6]. The signal is described as a convolution of a Gaussian scattering function with the width

$$B = \sqrt{\int_{path} f_B(t)\, dt} \quad \text{with} \quad f_B = \sigma N/R^2 \tag{1}$$

and the cosine response function of the grating interferometer. Here $\sigma$, $N$ and $R$ are described as the position dependent scattering cross section, particle density and correlation length of scattering structures in the sample, respectively. The integration is over the path of the radiation denoted with t. Note that the parameter $f_B$ corresponds to what was later called "linear diffusion coefficient" and assigned "$\varepsilon$" and "$\Omega$" for x-rays [26] and neutrons [29], respectively, while B was denoted "$\sigma$" for both. However, in contrast to (1) it is not further defined which sample features have influence on these parameters, but the exponential dependence of the DF signal on them, a consequence of the convolution with the modulation function is described. Besides the fact that a corresponding approximation is obviously not suited to extract quantitative small-angle-scattering (SAS) information, the fact that the signal is dependent on a product of the scattering power, involving thickness, concentration and scattering cross section, and a structural parameter implies that quantitative information is not accessible from a single measurement. Hence, all descriptions also in the latter attempts named above [27,28] are currently only suited to describe a priori well-known systems and systematic behavior related to thickness and eventually concentration etc. Another recent approach however, attempts to extract the scattering function by de-convolving the measured signal with the instrument function and thereby deriving the scattering function [25]. However, also such approach is limited concerning achieving a meaningful SAS function ready for quantitative analyses of structural parameters as is obvious when dealing with a cosine resolution function. Correspondingly, in [25] no claim is made to establish any quantitative relationship to structural sample parameters.

II. THEORY

As measurements with neutrons and x-rays are equivalent related to this technique, here no distinction shall be made between those.

Here an approach shall be introduced, which allows quantitative SAS information to be extracted from the dark-field signal measured in a grating interferometer yielding structural characterizations beyond direct spatial resolution of the imaging device without requiring to alter the sample. A solution can be found through a number of measurements with varied instrument parameters and by fully understanding the measured quantity. Here some previous work can provide hints already, especially Lynch et al [27] show the dependence of the dark-field signal on

$$\xi_{GI} = \frac{\lambda L_s}{p} \qquad (2)$$

which they call the autocorrelation length and where $\lambda$ is the wavelength, $L_s$ is the sample to detector distance [32,33] and $p$ is the period of the modulation function measured. They further derive a complex theoretical description for the signal of particles that can be described as hard spheres in solution. The description matches the measured signal for different particle sizes and measurements with different autocorrelation lengths reasonably well. Indeed these results already demonstrate, that not only the specific signal for one measurement can be calculated, but also that there is a systematic behavior of the signal dependent on the scattering structure, in this specific case the diameter of the spheric particles. The results clearly prove, that for this system scanning the autocorrelation length of the set-up allows for retrieving such structural information, though this conclusion is not explicitly provided. However, it seems obvious that scanning the correlation length is comparable to probing over a certain scattering vector range, i.e. q-range in a conventional SAS experiment.

In order to derive a general solution, a relation between conventional scattering and dark-field contrast imaging parameters can be found when describing the scattering angle $\theta$, for which in the SAS approximation $sin\theta \sim \theta \sim tan\theta$, as

$$\theta \sim \frac{x}{L_s} \text{ and } \theta \sim \frac{q\lambda}{2\pi}, \qquad (3)$$

respectively. Consequently one can express the scattering vector $q$ not only in terms of the scattering angle $\theta$ but also in terms of the position-shift $x$ of the interference pattern at a specific sample to detector distance $L_s$ in a grating measurement as

$$q = \frac{2\pi\theta}{\lambda} = \frac{2\pi x}{\lambda L_s} \tag{4}$$

The position shift $x$ corresponds to a phase shift $\Delta\omega$ of the modulation function of

$$\Delta\omega = \frac{2\pi x}{p} = \frac{2\pi L_s \theta}{p} = \frac{\lambda L_s}{p} q = \xi_{GI} q \tag{5}$$

where first $x$ is substituted by $L_s\theta$, then $2\pi\theta$ by $q\lambda$ and finally $\frac{\lambda L_s}{p}$ by $\xi_{GI}$. Having established a correlation between the induced phase shift $\omega$ of scattering to a specific angle $\theta$, the auto-correlation length and the modulus of the scattering vector $q$, it is important to understand how scattering and a specific scattering function impact the visibility measured with a grating interferometer set-up. The visibility $V$ is defined as $V=(I_{max}-I_{min})/(I_{max}+I_{min})$ of the measured intensity $I$, which is spatially modulated corresponding to the interference pattern introduced by the phase grating at a specific Talbot distance at which it is detected [4]. While radiation which is not scattered ($\theta_0=0$) and correspondingly arrives with $\Delta\omega = 0$ contributes at a maximum to the visibility $V$, as its particular visibility $V=V_0$, radiation which is scattered to a specific angle $\theta$, i.e. with a specific $q$ and hence a specific phase shift $\Delta\omega = \xi_{GI} q$ obviously contributes less. Consequently a degradation of the visibility is introduced, by scattered radiation producing a visibility $V<V_0$. Taking into account that the scattering function is symmetric around its centre at $q_0=0=\theta_0$, i.e. scattering to $\theta$ means equal scattering to $-\theta$ or $S(q) = S(-q)$, and with the basic relation

$$(\cos\Delta\omega + \cos(-\Delta\omega))/2 = \cos\Delta\omega \tag{6a}$$

that is valid for all points of the modulation function (Fig. 2) as

$$(\cos(\omega+\Delta\omega) + \cos(\omega-\Delta\omega))/2 = \cos\omega\,\cos\Delta\omega \tag{6b}$$

the corresponding visibility can be written as

$$V_S(\xi_{GI}, q) = V_0(\xi_{GI})\cos(\Delta\omega) = V_0(\xi_{GI})\cos(\xi_{GI} q) \,. \tag{7}$$

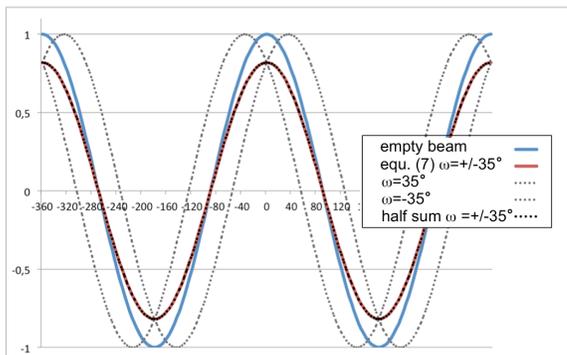

FIG. 2 Ideal cosine instrument response function for an empty beam measurement (blue line); instrument response function for scattering causing a phase shift of +/-35° calculated according to equ. (7) (red line)

and as the half sum of instrument response functions (black dots) of the empty beam shifted to 35° and -35° (gray dots), respectively.

Taking into account the whole scattering function $S(q)$ [34] and that the final visibility is a result of the convolution of the scattering function with the modulation function [6,26] of the grating interferometer this transforms into

$$V_S(\xi_{GI}) = V_0(\xi_{GI}) \int_{-Q_{max}}^{Q_{max}} dq\, S(q) \cos(\xi_{GI} q) \quad . \tag{8}$$

With regards to the fact that the scattering function $S(q)$ is a kind of Fourier transform of the real space correlation function

$$S(q) = \int_{-\infty}^{+\infty} d\xi\, G(\xi) \cos(\xi q) \tag{9}$$

and hence

$$G(\xi) = \int_{-\infty}^{+\infty} dq\, S(q) \cos(\xi q) \tag{10}$$

the normalized visibility can be given as

$$V_s(\xi_{GI}) / V_0(\xi_{GI}) = \int_{-Q_{max}}^{Q_{max}} dq\, S(q) \cos(\xi_{GI} q) = G(\xi_{GI}) \tag{11}$$

being directly proportional to the real space correlation function.

The grating or other modulation based dark-field SAS measurements [35] perform a back-transformation of the scattering function into real space and hence allow measuring directly the real space correlation function of a system.

However, the measured signal might also contain un-scattered neutrons. In order to account for that the macroscopic scattering cross section $\Sigma$ and the sample thickness t have to be taken into account and consequently

$$V_s(\xi_{GI}) / V_0(\xi_{GI}) = (1 - \Sigma t) + \Sigma t\, G(\xi_{GI}) \tag{12}$$

This situation and solution can be found equivalent to that of the well-known and well described spin-echo small-angle neutron scattering (SESANS)[36], where the response function can be described by a cosine dependence of the beam polarization, rather than a spatial function, on $\xi_{SE} q$ with $\xi_{SE}$ being the auto-correlation length of such set-ups referred to as spin-echo length [36]. It can be shown like it has been shown for SESANS that taking into account multiple scattering leads to [37]

$$\left.V_s(\xi_{GI})\middle/V_0(\xi_{GI})\right. = e^{\Sigma t(G(\xi_{GI})-1)}. \tag{13}$$

Obviously the simple multiplication with the sample thickness is only valid for a homogeneous sample, which one might assume in a SAS experiment but not so much for samples investigated in imaging, where such multiplication hence has to be replaced by the common integral along a specific path of the beam through the sample as

$$\left.V_s(\xi_{GI})\middle/V_0(\xi_{GI})\right. = e^{\int_{path} \Sigma(G(\xi_{GI})-1)dt} \tag{14}$$

with $\Sigma$ and $G$ being position dependent functions.

This establishes a complete description of the dark-field signal and its constitution, by replacing before used so-called material dependent constants referred to as linear diffusion coefficient [26,29] by well defined and established material parameters like the macroscopic scattering cross section and the real space correlation function. The latter finally provides the direct correlation of the signal with the structural parameters of the scattering structures, which is fundamental for any scattering method.

In addition, this solution implies that $\Sigma(G(\xi_{GI}) - 1)$ can even be reconstructed for a tomography for every correlation length $\xi_{GI}$ probed and hence the function $\Sigma(G(\xi_{GI}) - 1)$ can be retrieved for any position (x,y,z) in the sample corresponding to the spatial resolution of the set-up (voxel). This corresponds to a 3D resolved quantitative SAS measurement in case the tomography is performed for a number of correlation length values.

III. APPLICATION

In order to demonstrate the power and potential of this approach an example shall be given. Assuming a specimen of diluted hard spheres, for which the real space correlation function, like many others, is well known from SESANS [38] being

$$G(\zeta) = G(\xi/r) = \left[1 - \left(\frac{\zeta}{2}\right)^2\right]^{1/2}\left(1 + \frac{1}{8}\zeta^2\right) + \frac{1}{2}\zeta^2\left[1 - \left(\frac{\zeta}{4}\right)^2\right]ln\left[\frac{\zeta}{2+(4-\zeta^2)^{1/2}}\right] \tag{15}$$

and which has been stressed earlier [27,28] allows comparing the theory presented here with calculations and measurements presented in [27]. For that purpose the data presented in Fig. 2a (from Fig. 4 in [27]) is extracted and sorted by sample, i.e. different radii $r$ of spheric $SiO_2$ particles measured in a dispersion of $H_2O$, and by autocorrelation lengths used for the specific measurements like given in [27]. This data $\mu_d'(r,\xi_{GI})$ [27] is multiplied by the corresponding autocorrelation length $\xi_{GI}$ in order to get a correspondence with the function *2r(G($\xi_{GI}$)-1)* in the description derived here. Fig. 2a also demonstrates, that the calculation and data presented in [27] corresponds to *2r|G($\xi_{GI}$)-1|/$\xi_{GI}$*. In the presented theory the factor *r* is an integral part of the macroscopic scattering cross section $\Sigma$ for spheric particles which is defined as

$$\Sigma = (3/2)\phi_V \Delta\rho^2 \lambda^2 r . \tag{16}$$

Other parts of $\Sigma$ like the scattering length density contrast $\Delta\rho$, wavelength $\lambda$ and volume fraction $\phi_V$ on the other hand are not taken into account as they have been normalized with in [27] according to equ. d71 ibid. With these values a normalized visibility corresponding between the theory here and the data extracted from [27] is achieved with

$$\frac{V'_s}{V_0} = e^{-\mu_{d'}\xi_{GI}} = e^{2r(G(\xi_{GI})-1)} \tag{17}$$

and the results of both are plotted in Fig. 2b as a function of $\xi_{GI}$. A very good agreement is found, which proves that the theory very well describes the measurements. Furthermore, it is clearly visualized that important sample characteristics can not only be quantified, but easily be read from the data in particular for such kind of structure. The autocorrelation value at the saturation point, i.e. where the visibility does not decrease anymore with increasing autocorrelation length, directly provides the diameter of the hard spheres responsible for the scattering signal. This can already be seen from the transformation performed in Fig. 2a as compared to the representation in [27]. At $2r/\xi_{GI}\leq 1$, $|G(\xi_{GI}\geq 2r)-1|=1$. On the other hand it is obvious, that the visibility value of the saturation is directly related to the macroscopic scattering cross section $\Sigma$ of the system, i.e. to the volume fraction of the particles and the scattering length density contrast. That means at $\xi_{GI} = 2r$ the visibility value stabilizes at $V_s(\xi_{GI} \geq 2r)/V_0(\xi_{GI} \geq 2r) = e^{-\Sigma t}$. The information content of such measurement is equivalent to that of a conventional SAS measurement, however, spatial resolution is achieved without the requirement of scanning a pencil beam, like in conventional SAS experiments aiming for macroscopic 2D spatial resolution.

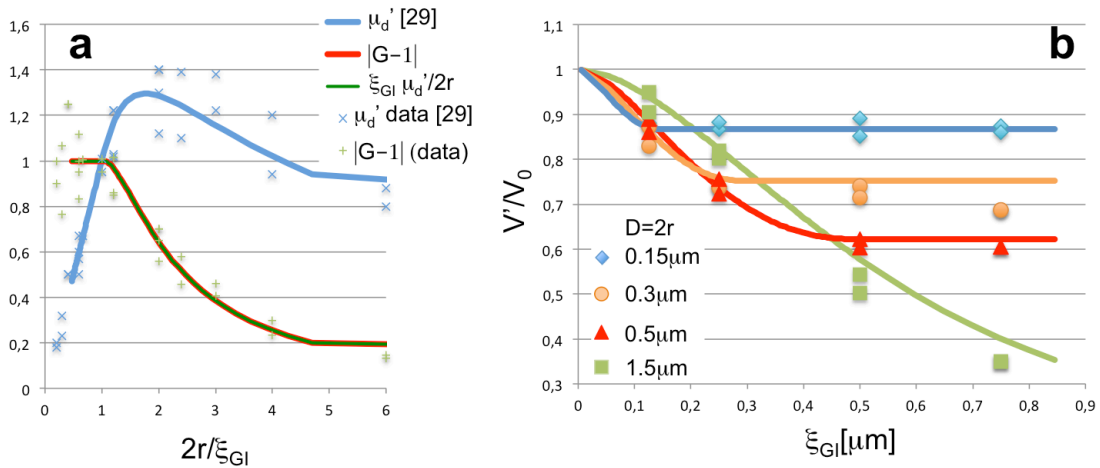

FIG. 3 (a) Representation of measured data $\mu_d$' and calculation as provided in [27] (blue line and symbols); Data and theory transformed according to $\xi_{GI} \mu_d'/2r$ (green line and symbols) and $|(G-1)|$ as calculated for the samples from the theory presented here (red line) displaying full agreement proving that the calculation presented and data reduction in [27] corresponds to $2r|(G-1)|/\xi_{GI}$;
(b) Visibility data (dots) from (a) [27] and calculation (lines) presented according to equ. (17) as a function of corresponding autocorrelation length $\xi_{GI}$ and sorted by sample (particle size), demonstrating good agreement between theory and data as well as the direct relation to particle size and scattering cross section (note: here represented by 2r only), both of which can be extracted straightforwardly.

IV. SUMMARY

In conclusion a quantitative and general relation between the measured visibility in grating interferometer based dark-field contrast imaging and the specific sample parameters of scattering cross section and in particular the real space correlation function of the structures in the sample that contribute to small-angle scattering and hence dark-field signal has been derived. It has been demonstrated how a scan of the autocorrelation length of the set-up in a dark-field contrast measurement yields the corresponding parameters and hence 2D SAS measurements become possible in such imaging mode providing full quantification of scattering parameters. The equivalence with a well known neutron scattering method, namely SESANS has been outlined, which in turn allows to build on the well know scattering functions, i.e. real space correlation functions, derived for this method in literature. Additionally, the presented theory underlines the potential of tomographic measurements, which allow gaining corresponding information with 3D spatial resolution. This theoretical assessment of the dark-field signal bears the potential to revolutionize the application of dark-field imaging, which already now as a qualitative tool is highly successful with x-rays as well as with neutrons. However, the knowledge of its quantitative character will without any doubt open numerous new fields in many areas of material science.